\documentclass[preprint]{aastex}  
\usepackage{psfig}

\begin{document}  

\title{Near-Infrared Synchrotron Emission from Cas A}  
\author{Jeonghee Rho }  
\affil{SIRTF Science Center, California Institute of Technology, MS 220-6,  
Pasadena, CA 91125; e-mail: rho@ipac.caltech.edu}  
\author{Stephen P.~Reynolds\footnote{Permanent address:
Department of Physics, North Carolina State
University, Raleigh, NC 27695}}
\affil{Harvard-Smithsonian Center for Astrophysics, 60 Garden St., 
Cambridge, MA 02138}
\author{ William T. Reach, Tom H. Jarrett }  
\affil{SIRTF Science Center/IPAC, California Institute of Technology, MS 220-6,Pasadena, CA 91125}  
\author{Glenn E. Allen}
\affil{MIT Center for Space Research, 77 Massachusetts Avenue, NE 80-6029, MA 02139-4307}
\author{John C. Wilson\footnote{Present Address: 255 Astronomy Bldg, 
University of Virginia, 530 McCormick Rd, Charlottesville, VA 22903.}}
\affil{Space Sciences Building, Cornell University, Ithaca, NY 14853}

\begin{abstract}

Recent high energy observations of Cas A suggests the presence of
synchrotron radiation, implying acceleration of cosmic rays by young
supernova remnants. We detect synchrotron emission from Cas A in the
near-infrared using Two Micron All Sky Survey (2MASS) and Palomar 200
inch PFIRCAM  observations.   The remnant is detected in J, H, and
K$_s$ bands using 2MASS: K$_s$ band is the brightest, H is moderate and
J is faint.  In the J  and H  bands,  bright [Fe II] lines (1.24$\mu$m
and 1.64$\mu$m) are detected spectroscopically.  The Palomar
observations include K$_s$ continuum, narrow-band 1.64 $\mu$m 
(centered on [Fe II]) and 2.12$\mu$m (centered on H$_2$(1-0)) images. 
While the narrow-band 1.64 $\mu$m image shows filamentary and knotty
structures, similar to the optical image,  the K$_s$ image shows a
relatively smooth, diffuse shell, remarkably   similar to the radio
image. The  H$_2$ image is identical to the K$_s$ continuum image with
surface brightness reduced  as expected for the ratio of filter
bandwidths, showing no contribution of H$_2$ lines to the K$_s$-band
image. The broad-band near-infrared fluxes of Cas A are generally
consistent with, but a few tens of percent higher  than, an 
extrapolation of the radio fluxes.   The hardening to higher
frequencies is possibly  due to  nonlinear shock acceleration  and/or
spectral index variation across the remnant.  We show evidence of
spectral index variation across  Cas A using  the `spectral tomography'
technique.  The presence of near-infrared synchrotron radiation
requires  the roll-off frequency   to be higher than 1.5 $\times$
10$^{14}$ Hz,  implying that  electrons  are accelerated to  energies
of at least  $E = 0.3 B_{\rm mG}^{-1/2}$ erg or 0.2 TeV. The
morphological similarity in diffuse emission between the radio and
K$_s$ band images implies that synchrotron losses are  not dominant, or
we would expect to see a greater concentration in knots.  We also show
dust continuum is not significant  in the near-infrared emission of Cas
A.    Our observations show unambiguous evidence that the near-infrared
K$_s$ band emission of Cas A is from synchrotron emission by
accelerated cosmic-ray electrons.

\end{abstract}

\noindent {\it Subject headings:} Supernova remnants: Cas A -- Near-infrared --Particle acceleration 

\section{Introduction}

The origin of cosmic rays has been unclear since their  discovery 
(Elster et al. 1969). It has been widely thought that their energy is
derived from supernovae, which are naturally associated with particle
acceleration because they are the primary energy input for interstellar
gas. Shocks are thought to be primary acceleration sites of cosmic
rays.  Some evidence that cosmic-ray electrons are accelerated in 
supernova remnant shocks has been recently found using X-ray and
gamma-ray observations. SN 1006 is a notable example. Its X-ray
emission shows featureless spectra in the limb, strong evidence  that
the emission is synchrotron radiation (Koyama et al. 1995). The
detection of TeV gamma-rays, probably  cosmic-ray background photons
upscattered by $\sim 100$ TeV electrons,  strengthened the claim of
synchrotron emission in SN 1006 (Tanimori et al. 1998).  There are a
few other SNRs including G347.3-05 \citep{koyama97,slane99},  and
G266.2-1.2 \citep{slane01}, which show featureless X-ray spectra,
possibly due to  synchrotron emission.  Other examples of synchrotron
emission are less clear, complicated by mixture with thermal emission,
such as Cas A and RCW 86 (Allen et al. 1997; Borkowski et al. 2000; Rho
et al. 2002).

However, even if synchrotron radiation commonly contributes to the
X-ray emission of SNRs, it does so at a level below the extrapolation
from radio frequencies, by factors of 3 to 100 for 14 Galactic SNRs
\citep{reynolds99} and 11 SNRs in the LMC \citep{hendrick01}.  That is,
the electron distribution responsible for radio synchrotron emission
must steepen at energies below 100 TeV in all those 25 cases.  This is
problematic if SNRs are to accelerate cosmic rays up to the slight
steepening around 3000 TeV in the cosmic ray ion spectrum known as the `knee.'
While relativistic electrons may be subject to radiative losses during
the acceleration process, this should not be a problem for ions, so
perhaps in all those cases electron acceleration is limited by losses,
while ion acceleration invisibly extends to much higher energies.  
Alternatively only certain remnants (for instance,
much older and larger ones than in the \citet{reynolds99} study)
may produce the highest energy cosmic rays.  Another possibility is that a
few remnants, such as Cas A, have very much stronger magnetic fields,
increasing both the acceleration rate and the importance of losses so
that the required steepening in the electron spectrum can be due to 
losses while ion acceleration can extend to $10^{15}$ eV and beyond
\citep{ellison99}.  Perhaps Cas A is a `super-accelerator' of cosmic
rays.

While the observed X-rays from Cas A fall below the radio extrapolation
by a smaller factor than any other Galactic SNR in the Reynolds \&
Keohane sample, this of course does not demonstrate that a portion of
those X-rays are due to synchrotron emission.   Most of soft
X-rays (below 7 keV) are clearly thermal (e.g., Gotthelf et al. 2001)
but emission is seen up to energies $> 10$ keV (Allen et al. 1997) with 
the {\it Rossi X-ray Timing Explorer} (RXTE),
and to $\sim 100$ keV with OSSE on the {\it Gamma Ray Observatory} \citep{the96}.
  The latter study
shows a power-law photon index of $-3.1 \pm 0.4$. A spectrum this hard
to such high energies cannot be produced by synchrotron emission from
shock-accelerated electrons in any reasonable model.  The most likely
explanation is nonthermal bremsstrahlung (NTB) (see, for instance,
Asvarov et al. 1989; Laming 2001) from the lowest-energy nonthermal
electron population, the beginning of the distribution that extends up
to radio-emitting electrons at around 1 -- 10 GeV.  Evidence for at
least some X-ray synchrotron emission near shock fronts was reported by
\citet{gotthelf01}, but the total emission in the apparent
synchrotron component is a small fraction of the total flux in the 4 --
6 keV range, and \citet{bleeker01} cite observations from {\it
XMM-Newton}   to assert only a small role for synchrotron emission in
the total X-ray flux.  Thus the question of the highest frequency at
which Cas A shows unambiguous synchrotron emission remains open.

Recently, a $5 \sigma$ detection of Cas A has been reported at
$\gamma$-ray energies above 1 TeV with the HEGRA Stereoscopic  
{\v C}erenkov Telescope System (Aharonian et al. 2001).  The emission could
be due to leptonic processes (NTB or inverse-Compton upscattering of
cosmic microwave background photons) or hadronic (inelastic cosmic-ray
proton collisions with ambient gas producing neutral pions which decay
to gamma-rays). NTB would require electrons of energies of a few TeV;
$\pi^0$-decay gamma rays would require similar energies for protons.
Inverse-Compton emission from the cosmic microwave background, however,
would indicate the presence of electrons of energies of order 100 TeV. 
However, until this detection is confirmed and the nature of the
processes producing the gamma rays clarified, the highest particle
energies present in Cas A are still uncertain.

The EGRET limits on emission above 100 MeV (Esposito et al. 1996) can be used
to put a lower limit on the magnetic-field strength of at least 0.4 mG
(Atoyan et al. 2000); if the field were weaker, the required
relativistic electron density would have to be higher to explain the
radio emission, and those electrons would emit NTB at levels higher
than the upper limit in the EGRET band. Thus Cas A seems to have a much
larger magnetic field than inferred for other remnants (e.g., 
$\sim 10\ \mu$G for SN 1006; Tanimori et al. 1998, Dyer et al. 2001).

The total flux of Cas A at millimeter wavelengths (Liszt \& Lucas 1999;
Mezger et al. 1986) lies very close to the extrapolation from
centimeter wavelengths.  Shortward of 1.2 mm, however, there is very
little information. Synchrotron continuum is required to fit the
mid-infrared ISOCAM and ISO-SWS (6-30$\mu$m) spectra  and the spectral
shape of synchrotron emission  (i.e. power law) dominates between
6--8$\mu$m (Douvion, Lagage, \& Pantin 2001). \citet{tuffs97} suspected
from ISO data that emission in a line-free region near 6 $\mu$m could
be synchrotron. Gerardy \& Fesen (2001) report that a K-band image
resembles the radio, but gave no quantitative flux determinations. 
Here we search for synchrotron emission from Cas A in the infrared, and
show a definitive detection of synchrotron radiation at near-infrared
wavelengths.  We discuss the implications of our result for theories of
particle acceleration and cosmic-ray origin in supernova remnants.

\section{Observations and Results}

\subsection{2MASS }

The 2MASS images of Cas A were obtained as part of the routine
observations  for  the Two Micron All Sky Survey (2MASS; Skrutskie et
al. 1997;  Cutri et al.
2002) using the northern
telescope at Mt. Hopkins, Arizona,  on 1998 December 26. The wavelength
coverages of the bands are   1.11-1.36$\mu$m (J), 1.5-1.8$\mu$m (H) and
2-2.32$\mu$m (K$_s$). The spatial resolution of 2MASS is 3$''$. The
detailed calibration of diffuse emission is explained in Rho et al.
(2001). Here we used updated  zero-point fluxes of 1592, 1024, and
668.8 Jy for J, H, and K$_s$, respectively.  Figure ~\ref{casa2mass}
shows the 2MASS three-color image of Cas A, revealing a shell-like
morphology for the near-infrared emission.  The K$_s$ and H band images
are brighter than the J band image, and the J band emission is barely
detected in 2MASS. The differences in morphology among the three bands
are not very noticeable except in the southwestern shell. The three
color image in Figure ~\ref{casa2mass} contrasts the difference: K$_s$
emission (reddish in color) dominates for most of Cas A, and H band
emission (greenish in color) is more noticeable in the southwestern
shell, some inner part of the northern shell, and sparsely distributed
knots.

We  obtained a few near-infrared spectra of Cas A using the CorMASS
instrument \citep{wilson} on the Palomar 60$''$ telescope.
The observations took place on Sep. 10-12 and Nov. 14-15, 2000.  
The CorMASS spectra towards the northwestern and southwestern shell 
show bright lines of [Fe~II] in the J and H bands (at 1.26$\mu$m and
1.64$\mu$m),  which is consistent with the spectra of 
\citet{gerardy01}, and no  lines in the K$_s$ band were detected. We
also took the spectra towards possibly synchrotron dominated positions
(positions D and C in Hughes et al. (2000) where no or weak X-ray
emission lines are detected),  and [Fe II] was not
detected. This indicates the line emitting region is spatially limited
across the remnant. 
The line contribution to
each filter is estimated based on the near-infrared spectra from 
\citet{gerardy01}, which 
is described in the next section. 

\subsection{Palomar 200 inch PFIRCAM images}

We observed Cas A on July 12--13, 2001 using the Prime Focus Infrared
Camera (PFIRCAM) on the Hale 200-inch telescope on Palomar Mountain.
The PFIRCAM has a $256\times 256$ pixel array, with a pixel scale of
0.494$''$ at the f/3.3 prime focus of the 200-inch telescope.   We took
K$_s$ broad band    ($\Delta\lambda=0.32\mu$m), [\ion{Fe}{2}] narrow filter
($\lambda=1.64\mu$m; $\Delta\lambda=0.016\mu$m)  and H$_2$ 
($\lambda=2.12\mu$m, $\Delta\lambda=0.02\mu$m) images.  Our goal
was to distinguish the structures between the line and continuum as
well as to measure the remnant fluxes. The seeing was 1$''$ to 1.5$''$.
We used the observational technique
described in Reach et al. (2002). 

The integration times per sky pixel for the K$_s$, \ion{Fe}{2}, and
H$_2$ images are 30, 60, and 60 sec, respectively. The PFIRCAM images
were calibrated by comparing photometry of sources in the field to the
2MASS catalog (Cutri et al. 2002).  The K$_s$ image was calibrated
using 68 stars with 2MASS K$_s$ magnitudes  ranging from 12.5 to 14.5,
the H$_2$-band image was calibrated using 43 stars with 2MASS K$_s$
magnitudes ranging from 11.0 to 14.5, and the [\ion{Fe}{2}] image was
calibrated using  52 stars with 2MASS H-band magnitudes ranging from
12.6 to 14.6.  Bright stars were excluded from the fit because of
non-linearity in the PFIRCAM data; the diffuse emission is equivalent to
stars fainter than 18$^{th}$ magnitude and is well within the linear
regime of the detector. The total remnant fluxes were measured from the
near-infrared images after subtracting stars  using the DAOPHOT package. The
regions (9$''$ radius) of two bright stars in the west were excluded from
the estimation of the total fluxes. 

The narrow-band mosaiced image of [\ion{Fe}{2}] in Figure
~\ref{pfirmcam} (in blue) is  globally similar to the optical image
\citep{douvion01,fesen87}, showing sharp filamentary and knotty
structures of Fe-rich ejecta.  However, while [Ne] and [\ion{S}{2}]
optical maps show brighter emission in the north, the [\ion{Fe}{2}]
image shows much brighter emission in the southwest.

Figures ~\ref{pfirmcam} (in red) and ~\ref{kbandradioeach}a show the 
Palomar PFIRCAM K$_s$ band image, which is noticeably different from
the \ion{Fe}{2}-band image. The K$_s$-band structure is much smoother
than the ejecta structure that dominates the [\ion{Fe}{2}] and optical
images.  The H$_2$-band image is essentially identical to the K$_s$
image, with surface brightness reduced (and noise increased) as
expected for the ratio of filter bandwidths; thus the H$_2$ band
appears to be completely due to the same continuum that generates the
the K$_s$-band image.  We show the 1.4 GHz radio image
\citep{anderson91} in Figure ~\ref{kbandradioeach}b  for comparison 
with the K$_s$ band image, because they are remarkably similar.  The
radio image of \citep{anderson91} has a pixel size of 0.4$''$ and a
spatial resolution of  1.3$''$, which is comparable to our Palomar
image.  The similarity between  the K$_s$ band and radio images
suggests that both the K$_s$ and radio images are due to the same
physical process: synchrotron radiation.

\section{Infrared Flux of Cas A}

\subsection{Total flux from near-infrared images}

The total remnant flux from the PFIRCAM K$_s$ band image is
0.32$\pm0.03$ Jy,  where the error includes the uncertainty in
calibration. In comparison, the 2MASS flux in the K$_s$ band is
0.31$\pm$0.02 Jy. Although the 2MASS diffuse emission may include some
faint stars due to lower  spatial resolution, the K$_s$ band flux is   
consistent between PFIRCAM and 2MASS. The statistical errors of the
measurement are less than 1\%, but a larger error is induced by the
method of star subtraction because sometimes the sampling of the fluxes
for the stars is insufficient or it is unclear if the bright blobs are
faint stars or diffuse emission, especially in the [\ion{Fe}{2}] image.
We experimented with different thresholds of star identification, and
we believe the error would not be larger than 20\% of the
measured flux (a conservative
estimate). Extinction varies across the remnant,  with A$_v$ ranging
typically from 4.3 to 6.2 mag \citep{searle71,fesen87}, and as high as
8 mag in the western boundary \citep{hurford96}.   The
extinction-corrected K$_s$ flux is  0.53$^{+0.12}_{-0.04}$ Jy, using
the near-infrared extinction law of \citet{rieke85}; the error includes
the possible range of extinction.  The CorMASS spectra show no lines in
the K$_s$ band, and  \citet{gerardy01} also detected  no lines within
the K$_s$ band (2-2.3 $\mu$m) from FMK (fast moving knot) positions in
the north and northeast of the remnant, while they detected faint lines
of  [Fe II], He I, and H II Br$\gamma$  from  QSF (quasi-stationary
flocculi) positions in the southwest.  It is likely the line
contribution is very small in the K$_s$ band image, considering that
(1) QSF covers a small portion of the remnant, (2) the morphology of
the K$_s$ image is remarkably similar to the radio image, and (3) the
narrow filter 2.12$\mu$m image is very similar to the K$_s$-band
image. 

The total flux of the PFIRCAM [\ion{Fe}{2}] image is 0.37$\pm$0.06 Jy, and  the
extinction-corrected flux is 0.76 Jy using A$_v$ of 4.5 mag.  This
total flux is  somewhat uncertain, because  the [\ion{Fe}{2}] emission is
knotty and filamentary and some of the bright peaks of  diffuse
emission may have been subtracted during subtraction of stars.   The 
2MASS total flux in the H band is  0.27 Jy, and the extinction-corrected
flux is 0.56 Jy. The 2MASS H band flux  may be somewhat overestimated
due to contribution from unresolved faint stars. Experimenting with
different star-subtraction methods shows that neither of these effects
is significant.

Emission lines contribute a small fraction of the flux observed in the
wide K$_s$ band. We estimate the line contribution as follows. 
We assume that the the emission in each band arises from a  combination
of a featureless continuum and spectral lines with a total brightness
in the H-band  that is a factor $f_L$ brighter than the [\ion{Fe}{2}]
1.64$\mu$m line. Using the observed fluxes and the filter widths,  the
line contribution to the H band is $\sim 2 f_L$\%, and even in the
narrow  [\ion{Fe}{2}]-centered filter, is only $\sim 30$\% due to the
1.64$\mu$m line. Spectroscopically, the 1.64 $\mu$m line is the
dominant line in the H band as found in observations by
\citet{gerardy01} from FMK positions.  While the 1.64 $\mu$m line is 
dominant ($f_L\simeq 1$) from FMK positions,  the QSF have a wide range
of lines, with $f_L\sim 2$. Thus the H band is dominated by continuum
emission, with lines contributing less than 4\% of the in-band flux. In
the K$_s$ band, the conclusion is even more clear, because of the
paucity of spectral lines. For most positions, there are no spectral
lines detected in the K$_s$ band  (in our CorMASS data and the FMK
positions from Gerardy and Fesen [2001]). Even using the line-rich QSF
spectra, the total line  brightness in the K$_s$ band is 10\% of the
total line brightness in the  H band, and spectral lines would
contribute $< 1$\% of the K$_s$-band flux. The conclusion that
continuum dominates the wide bands is robust, and even after accounting
for the uncertainties from star-subtraction methods, the line
contribution would be less than 8\% in the H bands, and less than 2\% in
K$_s$ band.

Measuring the J band flux was challenging, because 2MASS barely
detected the diffuse emission from the remnant. The 2MASS J band flux
is 0.12 Jy, and the extinction-corrected flux is 0.39 Jy using A$_v$ of
4.5 mag, which may be highly uncertain.  \citet{gerardy01} show a
higher resolution J band image, which is similar to our [\ion{Fe}{2}] 
band image. Their J band spectra show a number of bright lines such as
[\ion{Fe}{2}], [\ion{S}{2}] and [\ion{P}{2}] such that  the total flux
of lines in the J band is 7-9 times greater than in H-band. Using the
line fraction derived above for the H-band,  we estimate that that
lines contribute $\sim 20-40$\% of the J-band flux. The reason the
J-band and the narrow Fe-band images are similar is that they contain
similar fractions of line emission. 

\subsection{Possible dust contribution to Cas A flux}

The mid-infrared spectrum of Cas A revealed dust newly formed from
the ejecta material, as well as gas lines (Douvion, Lagage \& Pantin 2001; DLP01). 
A model for the mid-infrared spectrum using dust composed of pyroxene,
quartz, and aluminum oxide, with cold (90 K) and warm (350 K)
components can reasonably approximate the spectrum longward of 8 $\mu$m
(DLP01). We estimated the dust contribution to the near-infrared
brightness by calculating the ratio of the 2.2 $\mu$m to 8 $\mu$m
brightness of  small pyroxene (Dorschner et al. 1995) particles at 350
K,  then scaling to the observed 8 $\mu$m brightness (DLP01). In Figure
3 of DLP01 it is evident that dust dominates at 8 $\mu$m and above, and
the synchrotron component only begins to contribute  significantly at 6
$\mu$m. We find that the synchrotron emission is 3-4 orders of
magnitude higher than the  extrapolated dust model at 2.2 $\mu$m. We
also estimated the brightness of starlight scattered by the Cas A dust,
using the interstellar radiation field (Mathis, Mezger, and Panagia
1983), an albedo of 0.6, and the same optical depth as inferred from
the  mid-infrared thermal emission model (mostly from the colder
component). The scattering contribution is 4-5 orders of magnitude 
smaller than that of synchrotron emission. Therefore, at 2.2 $\mu$m,
synchrotron emission dominates over dust emission and
scattering.  Recent submillimeter  observations of Cas A show no dust
is present at 850 $\mu$m (Loinard 2002), which strengthens our
conclusion of no dust contribution at 2 $\mu$m.

\section{Near-Infrared Synchrotron Radiation from Cas A}
We show the broad-band spectrum of Cas A, from radio to TeV energies,
in Figure ~\ref{sed}. The data from Baars et al. (1977) for 10--35 MHz,
Liszt \& Lucas (1999) for 86 and 140 GHz, and \citet{mezger86} for
1.2mm data, are marked.  The high energy data of RXTE \citep{allen97},
OSSE \citep{the96}, EGRET \citep{esposito96}, and Whipple
\citep{lessard95} are also indicated.  Our near-infrared points are
above the fit to the radio data (solid line) with $\alpha$ = --0.77
(where log S= $\beta$ + $\alpha$ log $\nu$, $\beta$ = 5.745$\pm$0.025
for $\nu$ in MHz [Baars et al. 1977], which is equivalent to $\beta$
=10.365 for $\nu$ in Hz).  This shows that the K$_s$ flux before
correcting for extinction is comparable to the lower-frequency
extrapolation, and the extinction-corrected K$_s$ flux is $\sim$ 70 \%
higher than that extrapolation.  When we performed a weighted fit to
the radio, mm and near-IR data, we obtain an index that is not
significantly different from --0.77 and the value of the constant
$\beta$ is sightly higher  than that in Baars et al. (10.409$\pm$0.029,
see thick solid line in  Fig. ~\ref{sed}). The extrapolation of this
spectrum fell below the near-IR data points, suggesting positive
(concave-up) curvature. A flatter index than --0.77 has been suggested
by Mezger et al. (1986); they found --0.65.  When we fit the 86 and 140
GHz, 1.2mm and 2$\mu$m data alone, $\alpha$ = --0.709$\pm$0.040 and
$\beta$ = 9.762$\pm$0.444, showing a hardening to higher frequencies,
although we cannot completely rule out the possibility that the excess
near-IR flux above the extrapolation is due to contamination from line
emission.

It is possible that the concave spectrum between 140 GHz and  near-IR
is at least partially due to spectral-index variations across the
remnant.  If different shocks produce slightly different spectral
indices, perhaps because of differences in Mach number, shocks with
harder spectra will come to dominate at higher frequencies, and the
spatially integrated  spectrum will be concave.  Spectral-index
variations between 1.4 and 5 GHz were studied by \cite{anderson91}, who
obtained values from --0.64 to --0.92. Figure ~\ref{kbandradio} is a
two-color image  with the radio image (1.4 GHz; Anderson et al. 1991)
in red and the star-subtracted  K$_s$-band image in green, and shows 
spectral-index variations.  Radio-dominated regions (red in Figure
~\ref{kbandradio}) have a steeper spectral index, and K$_s$-band 
dominated regions (green in Figure ~\ref{kbandradio}) have a flatter
spectral index between the radio and K$_s$ frequencies.  The noticeable
steep spectral-index regions (red in Figure ~\ref{kbandradio}) are the
southeastern shell and knots, and a blob in the west outside of the
shell (at RA of 23$^{\rm h}$ 23$^{\rm m}$ 8.5$^{\rm s}$, $\delta$ =
58$^{\circ}$48$^{\prime}$44$''$).  By contrast, the southwest rim
appears to have a relatively flat spectrum. This image is indeed very
similar to the radio spectral-index map between 1.4 and 5 GHz,  which
is shown in Figure 3 of \cite{anderson91}.  Figure ~\ref{casaprofile}
shows  K$_s$-band and radio radial profiles averaged over $5^\circ --
10^\circ$ sectors toward the north, northeast, southeast, and southwest
directions (each position angle is given in the Figure caption). While
the K$_s$ band and radio profiles are similar in the N and NE (except
the radio is more sharply peaked in the shell),  the K$_s$ image is
noticeably fainter in the SE shell and brighter in the  SW shell, which
is consistent with what we see in Figure ~\ref{kbandradio}.

We have also constructed images using the `spectral tomography'
technique \citep{katzstone97}, in which scaled versions of one image
are subtracted from another. That is, for a series of trial values
$\alpha_t (< 0)$, we subtract from the radio image the K$_s$-band image
multiplied by $(\nu_{\rm radio}/\nu_{\rm IR})^{\alpha_t}$.  In such
images, emission with a spectrum steeper than $\alpha_t$ appears
positive, while flatter-spectrum emission appears negative.  In Figure
~7, we show images for $\alpha_t = -0.61, -0.67, -0.73,$ and $-0.77$.
Since the --0.61 image is essentially all positive, and the --0.77
image all negative, the radio-near IR spectral index of different
regions in Cas A is bracketed by these values. For $\alpha_t = -0.67$
and --0.73 we see both positive and negative structure. The
disappearance of much of the   emission along the bright southwestern
and northern rims in the --0.67 image indicates that the average 
radio-near IR spectral index in these regions is about --0.67. 
Similarly, the southeastern rim mostly vanishes in the --0.73 image 
shows that the average spectral index along the southeastern rim is
about that value.  These results are consistent with  the two-color
radio/K$_s$ image, in which the SE rim appears steeper, while most of
the rest of the remnant is flatter.   There is a great deal of other
small-scale structure in these images whose more detailed investigation
might provide useful information on the location and properties of
electron acceleration.

However, it is not clear that the observed integrated-spectrum
concavity can be produced by simply adding contributions from regions
with straight spectra with different slopes. For instance, a region
with a 1.4-5 GHz radio spectral index flatter than the average by 0.13
would be about 5 times brighter at K$_s$ band than a region of
comparable brightness at 1.4 GHz with the average spectrum.  Such large
variations do not appear to be the case.  But another explanation for
the concavity is available. Such concave-up curvature of the spectrum 
is predicted by nonlinear shock-acceleration models in which shock
transitions are broadened by pre-acceleration from cosmic rays
diffusing ahead of the shock, if particles' diffusion lengths increase
with energy.  In this case, more energetic particles feel a higher
velocity jump between upstream and downstream fluids, and the spectrum
is flatter at those energies \citep{eichler84, ellison91}.  Hints of
this effect are seen in Tycho's and Kepler's SNRs \citep{reynolds92} as
well as in Cas A.   If such concavity can be shown to be true for
localized regions as well as for the spatially integrated spectrum, it
will strongly support the case for such nonlinear effects. Other
theories of particle acceleration, such as stochastic (second-order
Fermi) acceleration in magnetohydrodynamic turbulence, predict a
steepening spectrum (e.g., Cowsik \& Sarkar 1984), so that our
observation supports diffusive (first-order Fermi) shock acceleration
instead.
A recent paper by Jones et al. (2003) also 
showed concavity of the spectrum from a small portion of the bright shell
with the  detection of polarized flux at 2.2$\mu$m, which 
reinforces presence of  concavity of the spectrum, i.e. presence
of non-linear shocks in Cas A.

The synchrotron spectrum should eventually roll off due to one of
several limitations on electron acceleration:  radiative losses, 
finite acceleration time, or particle escape (Reynolds 1998; Sturner et
al. 1997; Baring et al. 1999).  For any of these, the electron
distribution is likely to cut off no faster than exponentially
(Reynolds 1998) with an e-folding energy $E_{\rm max}$. For a
homogeneous source, the rolloff frequency is related to $E_{\rm max}$
by 
$$ {\nu_{\rm rolloff}} 
 =4.7 \times 10^{15}\,\, \biggl({{B_\perp\over{\rm mG}}}\biggr) 
      \biggl({{E_{\rm max}\over{\rm TeV}}}\biggr)^2 \,\, {\rm Hz}  $$
where $B_\perp$ is the sky-plane component of the magnetic field. A
lower limit for the roll-off frequency can be estimated  directly from
the high-frequency end of the K$_s$ band, which is 1.5 $\times$
10$^{14}$ Hz. In a magnetic field $B$, synchrotron radiation at this
frequency is produced by electrons with energies $E = 0.29 B_{\rm
mG}^{-1/2}$ erg or $\sim$0.2 TeV, where $B_{\rm mG}$ is the mean
magnetic field in milliGauss.  These values assume that $B$ lies
entirely in the plane of the sky,  since this gives us the most
conservative lower limit on $E$. Our claim of the presence of
electrons at these energies makes it plausible that some
portion of the X-ray emission is synchrotron as well.

Reynolds \& Keohane (1999) obtained an upper limit to the rolloff
frequency of 3.2$\times 10^{17}$ Hz, by assuming all the X-ray
emission to be synchrotron. Since the X-ray emission obviously
contains thermal emission with a rich line spectrum, this is a very
conservative upper limit.  Allen et al. (1997) used a cut-off
frequency of 2.4$\times 10^{17}$ Hz (1 keV). However, the cut-off
frequency is very sensitive to the values of spectral index from the
radio.

The half-life of electrons with Lorentz factor $\gamma$ radiating
in a magnetic field $B$ is given by
$$ t_{1/2} = {{ 5.1\times 10^{8} }\over {\gamma B^2}} \ {\rm s} 
 = { {1.81 \times 10^{16} } \, {\nu^{-1/2} \,
B_{mG}^{-3/2} } } \ {\rm s}$$
where we have averaged over magnetic-field directions.
The electrons radiating in the K$_s$ band ($\nu = 1.5 \times 10^{14}$
Hz) have lifetimes of only about 47 $B_{\rm mG}^{-3/2}$ yr.  With
typical velocities in Cas A of several thousand km s$^{-1}$, electrons
in such strong fields could travel distances of order tens of arcsec
(for $d = 3.4$ kpc; Reed et al. 1995) in a loss time. 
 The synchrotron emissivity varies roughly as
$n_{\rm els} B^{1 + \alpha}$, where $n_{\rm els}$ is the energy density
in relativistic electrons and $\alpha$ is the spectral index. If the
magnetic field evolves by flux freezing and the relativistic electrons
at these energies have relatively short diffusion lengths, we have
$n_{\rm els}$ is proportional to n, the thermal gas density, and $B$
proportional to a power of $n$ between 0 and 1.8.  Then the synchrotron
emissivity varies as roughly $n^2$, like the optical emissivity; both
optical emission and synchrotron emission would be prominant in high 
density regions, showing the morphological difference between the
radio and optical is not due to density difference. 
The smoothness of the K$_s$ band and radio images, relative to optical
images that more  directly trace the shock fronts and high thermal-gas
densities,  is due to the relatively large diffusion distances of the
extremely high-energy synchrotron emitting electrons, resulting in a
generally  smoother distribution of those electrons than that  of
thermal gas.

Even though the synchrotron loss time for the $K_s$-band-emitting
electrons is substantially less than the age of Cas A, it is much
longer than the time required for shock acceleration to produce
the electrons.  We can estimate the acceleration time based on
the general expression in Forman \& Morfill (1979), 
for the particular case of electron mean free path proportional
to gyroradius, $\lambda_{\rm mfp} = \eta r_g$ with $r_g = E/eB$ for
extreme-relativistic electrons, and making other reasonable 
assumptions detailed in Reynolds (1998). The acceleration time
for parallel shocks (shock normal parallel to the upstream magnetic
field) is

$$\tau_{\rm acc} = 1.3 \ \eta \left( B_{\rm mG} \right)^{-1} u_8^{-2} E 
                   \ {\rm yr}$$
                   
where $u_8$ is the shock speed in units of $10^8$
cm s$^{-1}$, and $E$ is in erg.
For perpendicular shocks, in which we assume a compression ratio of
4, the acceleration time is
$$\tau_{\rm acc} = 0.53 \ \eta^{-1} \left( B_{\rm mG} \right)^{-1} u_8^{-2} E 
                   \ {\rm yr.}$$

We infer that shocks can produce the
K$_s$-band emitting electrons in times of order one year.  This very
short timescale suggests that variability in acceleration, if not
depletion, of these electrons could be observed over timescales of the
same order. Now the radio morphology of Cas A makes it clear that there
is no single acceleration site like the outer shock wave in SN 1006;
instead, particle acceleration appears to take place in many small
knots all over the remnant \citep{anderson91}.  This picture can be
supported by detailed theoretical analysis (Atoyan et al. 2000).  In
their model, the magnetic field in the knots where acceleration takes
place is much larger than that in the general remnant volume, so one
would expect longer synchrotron lifetimes once electrons diffused out
of the acceleration zones.  In any case, to the extent that our $K_s$
band image resembles the radio, synchrotron losses are evidently not
dominant, or we would expect to see less diffuse emission and a greater
concentration in knots.  But small differences between the images of 
Figure \ref{kbandradio} shows that it is not impossible that this 
greater concentration has already begun to occur.

While most authors agree that shock acceleration produces the
relativistic electrons in Cas A, it is not impossible that some form of
turbulent or stochastic acceleration is responsible (e.g., Cowsik \&
Sarkar 1984).  One must ask if this mechanism, designed to describe
radio emission, can in the time available produce the $\sim 1$ erg
electrons required by our near-infrared image.  Melrose (1974) gives expressions
for acceleration  rates in magnetohydrodynamic turbulence which show
that once electrons are substantially suprathermal (`injected;' a
necessity for shock acceleration as well), they can reach relativistic
energies in a fraction of a year.  Subsequent acceleration of
relativistic electrons in MHD turbulence with amplitude $\delta B/B
\equiv \epsilon$ takes place on  a timescale
$$
\tau_{\rm acc} \sim {1 \over \epsilon^2} {r_g \over c} 
   \left({v_A \over c} \right)^{-2}$$
where $v_A$ is the Alfv\'en speed and $r_g$ is the electron's
gyroradius.  Strictly speaking, this result is true in the quasilinear
approximation ($\epsilon \ll 1$), but it should provide at least a
rough estimate for $\epsilon = 1$ to give us a conservative idea of the
fastest possible acceleration. In the above expression, $\epsilon$
refers to the amplitude of waves resonant with the particular particle,
which must be  presumed to exist (perhaps generated by the particles
themselves). Again assuming that the electron mean free
path is a multiple of its gyroradius, $\lambda_{\rm mfp}  \equiv \eta
r_g$, quasilinear theory then gives $\eta = (\delta B/B)^{-2} =
\epsilon^{-2}$, so we can rewrite the acceleration time as $\tau_{\rm
acc} \sim (\lambda_{\rm mfp}/c) (v_A/c)^{-2}$.  Since in the
extreme-relativistic limit, $r_g = E/eB$, with $E$ the particle energy,
the acceleration time to a particular energy is just proportional to
that energy.
For the strong fields we anticipate in Cas A, $v_A = 2.2 \times 10^8 \
B_{\rm mG} (\mu n_H)^{-1/2}$ cm s$^{-1}$, where the matter density is
$\rho = \mu m_p n_H$.  So $v_A/c \sim 10^{-2}$, and with $r_g = 2
\times 10^{12} E B_{\rm mG}^{-1}$ cm, we estimate $\tau_{\rm acc} \sim
10^6$ s to reach energies adequate to produce infrared synchrotron
radiation (in these strong fields) -- comparable to the rate of shock 
acceleration. This can occur because the strong magnetic fields in 
Cas A can give Alfven speeds comparable to the current, decelerated
outer blast-wave speed. So stochastic acceleration can
certainly  proceed rapidly enough to explain our observations.  The
spectrum produced by stochastic acceleration will depend on  the
properties of the turbulence and it is not clear if the required
remarkably flat, slightly hardening spectrum can in fact be produced.
However, neither shock acceleration nor stochastic acceleration can
be ruled out on the basis of acceleration-time arguments.

\section{Summary and conclusions}
   
The 2MASS near-infrared images reveal emission from Cas A in the J, H,
and K$_s$ bands. The images reveal a shell-like morphology  shown in
Figure ~\ref{casa2mass}.  The K$_s$ and H band images are brighter than
the J band image. The three-color image  demonstrates the difference:
K$_s$ emission  is dominant for the majority of Cas A, and H band
emission is noticeable in the southwestern shell.   Near-infrared
spectra showed  [\ion{Fe}{2}] lines in J (1.25$\mu$m) and H
(1.64$\mu$m),  and no lines were detected within the K$_s$ band. 

A high-resolution Palomar PFIRCAM narrow-band image of [\ion{Fe}{2}]
(1.64$\mu$m) (see Figure ~\ref{pfirmcam}) is  similar to the optical
image  of Cas A, showing sharp filamentary and knotty structures of
Fe-rich ejecta.  The PFIRCAM K$_s$ band image  is much smoother and
more diffuse than the ejecta structure  that dominates the optical
images (and is present in the [\ion{Fe}{2}] image). Instead, the K$_s$
image is remarkably similar to the radio image of Cas A, suggesting
that the physical mechanism that produces radio emission--synchrotron
radiation--is also  responsible for the K$_s$ band emission.  We
measure a total flux of Cas A in the K$_s$ band of 0.32$\pm0.03$ Jy,
implying a flux of 0.53$^{+0.12}_{-0.04}$ Jy after correcting for
extinction.   The spectral energy distribution of Cas A shows that the
near-infrared flux is close to an extrapolation of the radio fluxes, 
supporting our interpretation of the near-IR flux as synchrotron
radiation. The near-infrared K$_s$ band fluxes are actually a few tens
of percent higher than an extrapolation of the radio fluxes. Such 
positive (concave-up) curvature of the spectrum (hardening to higher
frequencies) is predicted by nonlinear shock-acceleration models in
which shock transitions are broadened by pre-acceleration from cosmic
rays diffusing ahead of the shock, if particles' diffusion lengths
increase with energy.

The synchrotron spectrum may roll off due to any of several possible
limitations on electron acceleration including radiative losses,
acceleration time, or particle escape; this is  measured as the rolloff
frequency.   Our near-infrared emission measurement gives us a lower
limit to the roll-off frequency of 1.5 $\times$ 10$^{14}$ Hz.
Synchrotron radiation at this frequency is produced by electrons with
energies  $E = 0.3 B_{\rm mG}^{-1/2}$ erg or about 0.2 TeV, where
$B_{\rm mG}$ is the mean magnetic field in milliGauss.  The synchrotron
loss time for  electrons radiating in the K$_s$ band  is only about 47
$B_{\rm mG}^{-3/2}$ yr. The fact that such electrons can thus travel
only distances on the   order of tens of arcsec implies particle
acceleration takes place in many sites all over the remnant. Our $K_s$
band image resembles the radio, so synchrotron losses evidently do not
dominate the small-scale structure. The acceleration time scale for
electrons with energy high  enough to produce the near-infrared
emission from Cas A is of order a  year (or less), so either shock or
stochastic acceleration is rapid enough to accelerate electrons to the
required energies.

Our results confirm that supernova remnants are high-energy cosmic-ray
acceleration sites that can produce 0.2 TeV particles.  We also show
that near-infrared observations are a good tool to study synchrotron
radiation and cosmic-ray acceleration in supernova remnants.

\acknowledgements J. Rho thanks Pierre-Olivier Lagage for discussion on
synchrotron emission from mid-infrared observation, which gave
her strong motivation for this project. 

{} 
\clearpage

\begin{figure}
\caption{Mosaiced 2MASS three color Atlas Image of Cas A. The color
ranges are 0.7-3.7 $\mu$Jy (J in blue), 5-29 $\mu$Jy (H in green),  and
7-38 $\mu$Jy (K$_s$ in red) per 1$''$ pixel. Most of the diffuse
emission is K$_s$ band (red) dominated, but there are a few places
showing significant H band (green) emission, in particular, in the
southwestern shell. The position of the center is  $(23^{\rm h} \
23^{\rm m} \ 25.7^{\rm s}, \  58^{\circ} \ 48^{\prime}
52^{\prime\prime}$ in J2000), and the image  size is 12$'$.}
\label{casa2mass}
\end{figure}

\begin{figure}
\caption{Mosaiced Palomar PFRICAM two-color images of K$_s$ (red) and 
[\ion{Fe}{2}] (blue). The narrow-band [\ion{Fe}{2}] image shows 
filamentary and  knotty structures (from ejecta material) that are not
present in the broad-band K$_s$ image, which is dominated by
synchrotron emission. The colors are 5-82 $\mu$Jy (blue) and 2-11
$\mu$Jy (red) per 0.5$''$ pixel. The center of the image is the same as
Fig. 1 and the image size is 6.4$'$.}
\label{pfirmcam}
\end{figure}

\begin{figure}
\caption{(a) The K$_s$ band image of Cas A.  (b) Radio image of Cas A
at 1.4 GHz \citep{anderson91}. The K$_s$ and radio images are
remarkablely similar, except of course for the presence of unrelated
stars in the K$_s$ image.}
\label{kbandradioeach}
\end{figure}

\begin{figure}
\caption{Broad band frequency and flux  spectrum of Cas A.  The
near-infrared K$_s$ flux is marked with an asterisk and the extinction
corrected flux  is marked as a square (assuming A$_v$=4.5 mag).  The
solid line has a spectral index of --0.77, which is derived using all the
radio,  near-IR, 86 GHz, and 140 GHz fluxes, with an exponential  cut-off
applied. The dash-dotted line has a spectral index of --0.71, which
is derived using only the near-IR,
86 GHz, and 140 GHz fluxes. 
The dash-two-dotted lines are nonthermal-bremsstrahlung (NB), 
inverse Compton scattering
of the cosmic microwave background (IC), and the decay of neutral pions
($\pi^0$) from Allen et al. (1997). 
}
\label{sed}
\end{figure}

\begin{figure}
\caption{Two color images of K$_s$ band (green) and radio (red). Stars
have been subtracted from the K$_s$ band image. The area emitting in
both the K$_s$ band and radio appears in yellow. The three  saturated
stars are masked. This image highlights differences between the K$_s$
and radio emission from the remnant, with the southeast being
relatively radio-bright.}
\label{kbandradio}
\end{figure}

\begin{figure}
\caption{Radial profiles of K$_s$ band (solid) and radio (dashed)
images. We chose the center position of RA = 23$^{\rm h}$  23$^{\rm m}$
26$^{\rm s}$, Dec = 58$^{\circ}$48$^{\prime}$42$''$ (J2000);  N,
covering position angles (eastward from the north) of
355-360$^{\circ}$; NE, covering P.A. of 45-50$^{\circ}$;  SE, covering
P.A. of 135-145$^{\circ}$; SW, covering P.A. of 210-220$^{\circ}$. The
relative fluxes are conserved within each of K$_s$ and radio images. }
\label{casaprofile}
\end{figure}

\begin{figure}

\caption{`Spectral tomography' maps: from the radio image we subtract
the K$_s$-band image multiplied by $(\nu_{\rm radio}/\nu_{\rm
IR})^{\alpha_t}$. White corresponds to positive values, softer 
(steeper) than the trial index, and dark to negative values, harder 
than the trial index. The scale is between -0.03 to 0.03. The three
saturated stars are masked shown in white. 
 (a: top left) ${\alpha_t}$ = --0.61: the spectral
index is steeper than --0.61 for all of Cas A. (b: top right)
${\alpha_t}$ = --0.67: SW shell turns positive to  negative. (c: bottom
left) ${\alpha_t}$ = --0.73:  northern shell and outer shock of
northern shell change to negative. (d: bottom right) ${\alpha_t}$ =
--0.77: the only positive structure is the western edge knot.   We
deduce that the approximate  spectral index is $\sim$--0.67 for SW
shell, $\sim$--0.73 for  the northern shell and the upper part of SE
shell, $\sim$--0.77  for the lower part of SE shell, and steeper than
$\sim$--0.77 for the western edge knot.}
\label{casaspindex}
\end{figure}

\def\extra{
\clearpage
\vskip 2truecm
 {Fig. 1 is not included here (see separate file in jpg format)}
\vskip 2truecm
{Fig. 2  is not included here (see separate file in jpg format)}

\vskip 2truecm
{Fig. 3a and 3b  are not included here (see separate file in jpg format)}
\newpage
\psfig{figure=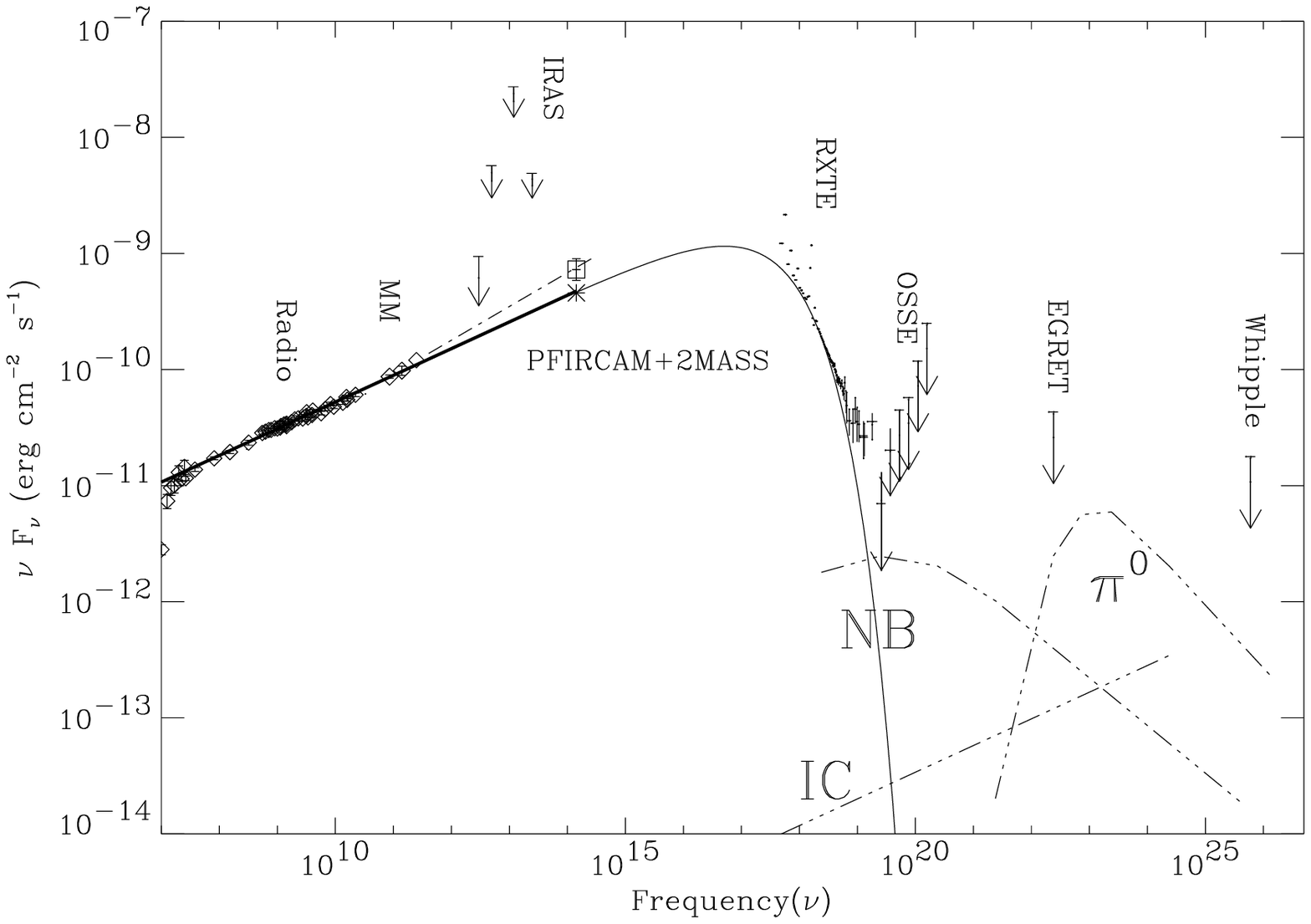}
{Fig. 4}

\newpage

\vskip 5truecm
{Fig. 5  is not included here (see separate file in jpg format)}

\newpage
\plotone{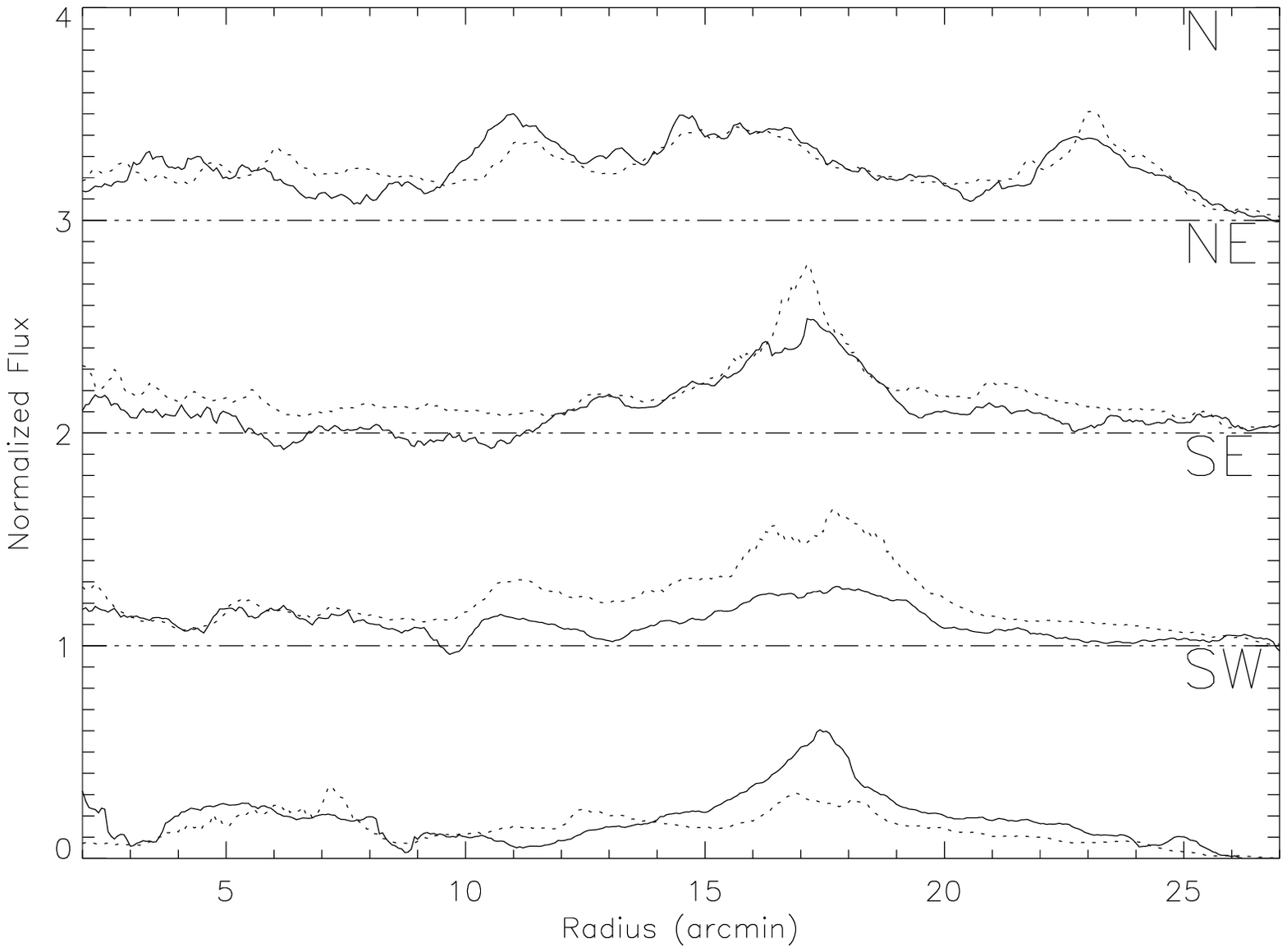}
{Fig. 6}

\psfig{figure=f7.ps,width=17truecm}

{Fig. 7}
}

\begin{thebibliography}{} 
\bibitem[Allen et al.(1997)]{allen97}
Allen, G.E., Keohane, J.W., Gotthelf, E.V., Petre, R., \& Johoda, K., 
1997, ApJL, 487, L97

\bibitem[Anderson et al.(1991)]{anderson91} Anderson, M. A.,  \& Rudnick,
L., Leppik, P., Perley, R., \& Braun, R., 1991, 373, 157

\bibitem[Aharonian et al.(2001)]{aharonian01}
Aharonian et al. 2001, A \&A, 370, 112

\bibitem[Asvarov et al.(1989)]{asvarov89}
Asvarov, A.I., Guseinov, O.Kh., Dogel, V.A., \& Kasumov, F.K.
1989, Sov.Astron. 33, 532

\bibitem[Atoyan et al.(2000)]{atoyan00}Atoyan, A.M., 
Tuffs, R.J., Aharonian, F.A., \& V\"olk, H.J. 2000,
A\&A, 354, 915

\bibitem[Baars et al.(1977)]{baars77}
Baars, J.W.M., Genzel, R., Pauliny-Toth, I.I.K., \& Witzel, A., 
1977, A\&, 61, 99

\bibitem[Baring et al.(1999)]{baring99}
Baring, M.G., Ellison, D.C., Reynolds, S.P, Grenier, I.A., \& Goret, P.,
1999, 513, 311

\bibitem[Bleeker et al.(2001)]{bleeker01} Bleeker, J.A.M., Willingale, R., 
van der Heyden, K., Donnerl, K.,
Kaastra, J.S., Aschenbach, B., \& Vink, J. 2001, A\&A, 
365, L225

\bibitem[Borkowski et al.(2001)]{brrd01}
Borkowski, K. J., Rho, J.,  Reynolds, S. P., \& Dyer, K. K. 2001, ApJ,
550, 334

\bibitem[Cowsik \& Sarkar(1984)]{cowsik84}
Cowsik, R., \& Sarkar, S. 1984, MNRAS, 207, 745 (erratum, 1984,
MNRAS, 209, 719)

\bibitem[Dorschner et al.(1995)]{dorschner95} Dorschner, J.,
Begemann, B., Henning, T., J\"ager, C., \& Mutschke, H. 1995,
\aap, 300, 503--520

\bibitem[Douvion, Lagage, \& Pantin(2001)]{douvion01}
Douvion, T.,  Lagage, P.O. \& Pantin, E., 2001, A\&A, 369, 589 (DLP01)

\bibitem[Dyer et al.(2001)]{dyer01}Dyer, K.K., Reynolds,
S.P., Borkowski, K.J., Allen, G.A., \& Petre, R. 2001,
ApJ, 550, 334

\bibitem[Eichler(1984)]{eichler84}Eichler, D. 1984, ApJ,
277, 429

\bibitem[Ellison et al.(1999)]{ellison99}
Ellison, D.C., Goret, P., Baring, M.G., Grenier, I.A., 
\& Lagage, P.-O. 1999, Proc. 26th Int. Cosmic Ray Conf. (Salt
Lake City), 3, 468

\bibitem[Ellison \& Reynolds(1991)]{ellison91}
Ellison, D.C., \& Reynolds, S.P. 1991, ApJ, 378, 214

\bibitem[Esposito et al.(1996)]{esposito96}
Esposito, J.A., Hunger, S.D., Kanbach, G., \& Sreekumar, P.,
1996, ApJ, 461, 820

\bibitem[Fesen, Becker \& Blair(1987)]{fesen87}
Fesen, R.A., Becker, R.H., \& Blair W.T., 1987, ApJ, 313, 378

\bibitem[Gerardy \& Fesen(2001)]{gerardy01}
Gerardy, C. L.,  \& Fesen, R. A.,  2001, ApJ, 121, 2781

\bibitem[Gotthelf et al.(2001)]{gotthelf01}
Gotthelf, E.V., Koralesky, B., Rudnick, L., Jones, T.W.,
Hwang, U., \& Petre, R. 2001, ApJ, 552, L39

\bibitem[Hendrick \& Reynolds(2001)]{hendrick01}
Hendrick, S.P., \& Reynolds, S.P. 2001, ApJ, 559, 903

\bibitem[Hurford \& Fesen(1996)]{hurford96}
Hurford, A. P. \& Fesen, R. A., 1996, ApJ, 469, 246

\bibitem[Jones(2002)]{jones02}Jones, T.J., Rudnick, L., Delaney, T., \& 
Bowden, J., 2003, ApJ, in press (also see astro-ph/0212544)

\bibitem[Katz-Stone \& Rudnick(1997)]{katzstone97}
Katz-Stone, D.M., \& Rudnick, L.   1997, ApJ, 488, 146
 
\bibitem[Koyama et al.(1995)]{koyama95}
Koyama, K., et al. 1995, 47, 711

\bibitem[Koyama et al.(1997)]{koyama97}
Koyama, K., et al. 1997, PASJ, 49, L7

\bibitem[Laming(2001)]{laming01}Laming, J.M. 2001, ApJ, 546, 1149

\bibitem[Lessard et al.(1995)]{lessard95}
Lessard, R.W., et al. 1995, in Proc. 24th Int. Cosmic-Ray conf. (Rome),
2, 475
\bibitem[Loinard(2002)]{loinard02}
Loinard, L., 2002, ``Winds, Bubbles and Explosions" Patzcuaro in Mexico,
in press
\bibitem[Liszt \& Lucas(1999)]{liszt99}
Liszt, H., \& Lucas, R., 1999, A\&A, 347, 258


\bibitem[Mathis, Mezger, \& Panagia(1983)]{mathis83}
Mathis, J.S., Mezger, P.G., \& Panagia, N. (1983), A\&A, 128, 212

\bibitem[Melrose(1974)]{melrose74}
Melrose, D. B., 1974,SoPh, 37, 353

\bibitem[Mezger et al.(1986)]{mezger86}
Mezger, P.G., Tuffs, R.J., Chini, R., Kreysa, E., \& Gem\"und, 1986,
A\&A, 167, 145

\bibitem[Reach et al.(2002)]{reach02}
Reach, W. T., Rho, J., Jarrett, T. H., \& Lagage, P.-O. 2002, 
ApJ, 564, 302

\bibitem[Reed et al.(1995)]{reed95}
Reed, J.E., Hester, J.J., Fabian, A.C., \& Winkler, P.F., 1995, ApJ,
440, 706

\bibitem[Reynolds(1998)]{reynolds98} Reynolds, S. P., 1998, ApJ, 493, 375

\bibitem[Reynolds \& Ellison(1992)]{reynolds92}
Reynolds, S.P., \& Ellison, D.C. 1992, ApJ, 399, L75

\bibitem[Reynolds \& Keohane(1999)]{reynolds99}
Reynolds, S.P., \& Keohane, J.W. 1999, ApJ, 525, 368

\bibitem[Rho et al.(2001)]{rho01}
Rho, J., Jarrett, T.H., Cutri, R.M., \& Reach, W.T., 2001, ApJ, 547, 885

\bibitem[Rho et al.(2002)]{rho02}
Rho, J., Dyer, K.K., Borkowski, K.J., \& Reynolds, S.P., 2002,
ApJ, 581, 1116

\bibitem[Rieke \& Lebofsky(1985)]{rieke85}
Rieke, G. H., \& Lebofsky, M.J., 1985, ApJ, 288, 618


\bibitem[Searle(1971)]{searle71} Searle, L., 1971, ApJ, 168, 41

\bibitem[Skrutski et al.(1997)]{bla} Skrutskie, et al., 1997, $``$The impact of large scale near-IR sky
surveys", ed.by F. Garzon et al., p25

\bibitem[Slane et al.(1999)]{slane99} Slane, P. , Gaensler, B.  M.,
Dame, T. M., Hughes, J. P., Plucinsky, P. P. \& Green, A.  1999, \apj,
525, 357

\bibitem[Slane et al.(2001)]{slane01} 
Slane, P., Hughes, J.P., Edgar, R.J., Plucinsky, P.P., Miyata, E., H. Tsunemi, \
\& B. Aschenbach., 2001, ApJ, 548, 814

\bibitem[Sturner et al.(1997)]{sturner97}
Sturner, S.J., Skibo, J. G., Dermer, C. D., \& Mattox, J.R., 1997, ApJ, 490, 619

\bibitem[Tanimori et al.(1998)]{tanimori98}
Tanimori, T., et al. 1998, ApJ, 497, L25

\bibitem[The et al.(1996)]{the96}
The, L.-S., Leising, M.D., Kurfess, J.D., Johnson, W.N.,
Hartmann, D.H., Gehrels, N., Grove, J.E., \& Purcell, W.R.
1996, A\&AS, 120, 357

\bibitem[Tuffs et al.(1997)]{tuffs97}
Tuffs, R.J., Drury, L., Fischera, J., et al.
1997, Proc. 1st ISO Workshop on Analytical Spectroscopy (ESA 
SP-419), p. 177

\bibitem[Wilson et al.(2001)]{wilson} Wilson, J. C. et al. 2001, \pasp, 113, 227

\end{thebibliography}
\end{document}